\documentclass[11pt]{article}
\pdfoutput=1
\usepackage{amsmath}
\usepackage{slashed}
\usepackage{jheppub}
\usepackage{subfig}

\newcommand\beq{\begin{equation}}
\newcommand\eeq{\end{equation}}
\newcommand\be{\begin{equation}}
\newcommand\ee{\end{equation}}
\def\be{\begin{eqnarray}}
\def\ee{\end{eqnarray}}

\def\Dslash{\,\,{\raise.15ex\hbox{/}\mkern-12mu D}}
\def\Dbarslash{\,\,{\raise.15ex\hbox{/}\mkern-12mu {\bar D}}}
\def\delslash{\,\,{\raise.15ex\hbox{/}\mkern-9mu \partial}}
\def\delbarslash{\,\,{\raise.15ex\hbox{/}\mkern-9mu {\bar\partial}}}
\def\pslash{\,\,{\raise.15ex\hbox{/}\mkern-9mu p}}
\def\calDslash{\,\,{\raise.15ex\hbox{/}\mkern-12mu {\cal D}}}

\newcommand{\Tr}{{\rm Tr}}

\def\lae{\mathrel{\mathop{\smash{\lower .5 ex \hbox{$\stackrel<\sim$}}}}}
\def\lae{\mathrel{\mathop{\smash{\lower .5 ex \hbox{$\stackrel>\sim$}}}}}

\title{Supersymmetric Adler Functions and Holography}

\author[a]{Masaya Iwanaga,}
\author[b]{Andreas Karch}
\affiliation[a]{Department of Physics, Nagoya University, Nagoya 464-8602, Japan}
\affiliation[b]{Department of Physics, University of Washington, Seattle, Wa, 98195-1560, USA}
\author[a,c]{and Tadakatsu Sakai}
\affiliation[c]{Kobayashi-Maskawa Institute for the Origin
of Particles and the Universe, Nagoya University, Nagoya 464-8602, Japan}

\preprint{\today}

\emailAdd{tokidokinemutai@gmail.com, akarch@uw.edu, tsakai@eken.phys.nagoya-u.ac.jp}

\abstract{We perform several tests on a recent proposal by Shifman and Stepanyantz for an exact expression for the current correlation functions in supersymmetric gauge theories. We clarify the meaning of the relation in superconformal theories. In particular we show that it automatically follows from known relations between the current correlation functions and anomalies. It therefore also automatically matches between different dual realizations of the same superconformal theory. We use holographic examples as well as calculations in free theories to show that the proposed relation fails in theories with mass terms.
}

\begin{document}
\maketitle


\section{Introduction}

In \cite{Shifman:2014cya,Shifman:2015doa} Shifman and Stepanyantz proposed an exact relation for the Adler function in $d=4$ dimensional ${\cal N}=1$ supersymmetric gauge theories. The Adler function in essence is the current/current correlations function for a global $U(1)$ current.
\beq
\label{dfun}
D(Q^2) \equiv - 12 \pi^2 (Q^2 d/dQ^2) \Pi(Q^2)
\eeq
where
\beq i \int d^4x \, e^{i q x}
 \langle T \{ j_{\mu}(x) j_{\nu}(0) \}
\rangle \equiv (q_{\mu} q_{\nu} - q^2 g_{\mu \nu} ) \Pi(Q^2)
\eeq
and $Q^2 = -q^2$, that is we denote with $q^2$ timelike momenta, whereas $Q^2$ denotes the analytic continuation to spacelike momentum.

The proposed relation for the Adler function in a gauge theory based on the gauge group $SU(N_c)$ is
\beq
\label{adler}
D(Q^2) =\frac{3}{2} N_c \sum_f q_f^2 [1 - \gamma(\alpha_s(Q^2)].
\eeq
Here the sum runs over all the matter fields in the theory, $q_f$ is the charge of the $f$-th matter field under the global $U(1)$ symmetry and $\gamma$ its anomalous dimension (which depends on the strong coupling constant at the scale $Q^2$). This relation is in spirit similar to the celebrated exact NSVZ \cite{Novikov:1983uc} expression for the $\beta$-function in ${\cal N}=1$ supersymmetric gauge theories, which asserts that
\beq
\label{beta}
\beta(g) = - \frac{g^3}{16 \pi^2} \frac{3 N_c - N_f + N_f \gamma(g^2)}{1 - N_c \frac{g^2}{8 \pi^2}}.
\eeq
In both cases the relation equates two objects which themselves receive corrections order by order in perturbation theory, but the relation between them is supposed to be exact to all orders and even non-perturbatively. Furthermore, both $\beta(g)$ and $\gamma(g)$ depend on a choice of scheme. But the proposal asserts that \eqref{adler} holds in the same scheme in which \eqref{beta} is true.

In this work, we would like to check this relation and analyze its consequences. First we look at superconformal field theories, where the relation is trivially satisfied. It was argued in \cite{Shifman:2015doa} that in this case the Adler function could be used to give additional tests of proposed dualities but what we find is that, given \eqref{adler}, agreement of the current/current correlation functions is already guaranteed by matching of 't Hooft anomalies. Furthermore, we calculate the Adler function exactly in non-conformal theories with holographic gravity dual in order to understand the role that \eqref{adler} plays in that context. In our example, conformal invariance is broken via a mass term. In the derivation of \eqref{adler} in \cite{Shifman:2015doa} the mass of the matter fields is set to zero from the get go. Our holographic example demonstrates that \eqref{adler} fails in these theories with massive matter. We confirm this conclusion by studying a free theory.

\section{Conformal Field Theories}

The definition \eqref{dfun} assures that the D-function is a constant in
conformal theories. In 4 spacetime dimensions, conformal invariance fixes the current/current correlation function to go as $x^{-6}$, and its Fourier transform than has to be proportional to  $\log(q^2)$. So the derivative just picks out the prefactor of the logarithm. Similarly, the anomalous dimensions are
just $q$-independent numbers at the fixed point.

In section 5 of \cite{Shifman:2015doa} Shifman and Stepanyantz make the interesting point that $D$ is a
physical observable, so it should agree at an IR fixed point in a pair
of dual theories and potentially could be used as an additional check of the Seiberg dualities \cite{Seiberg:1994pq}. What we will see here is that matching of the $D$ functions is already guaranteed by matching of the R-charge anomaly. So while not an independent confirmation of Seiberg duality, it is reassuring that the picture hangs nicely together. A crucial extra tool we have in superconformal theories is the fact that the superconformal algebra relates the R-charge $r$ of a chiral primary operator to its dimension $\Delta$. For a chiral superfield one has
\beq
\label{rdim}
\Delta \equiv 1 + \frac{\gamma}{2} = \frac{3}{2} r
\eeq
Let us, for example, look at supersymmetric QCD, that is a ${\cal N}=1$ supersymmetric gauge
theory based on an $SU(N_c)$ gauge group with $N_f$ flavors of fundamental representation chiral multiplets $Q$ and anti-fundamental representation chiral multiplets $\tilde{Q}$, the quarks. This theory is believed \cite{Seiberg:1994pq} to flow to a strongly interacting conformal field theory in the window $3/2 < N_f/N_c < 3$. In this conformal window the theory has a dual description in terms of a $SU(N_f-N_c)$ gauge theory with dual quarks $q$ and $\tilde{q}$ as well as a meson singlet chiral multiplet $M$ coupled to the dual quarks via a superpotential $W=M q \tilde{q}$.
The anomaly free R-charge assignments are
\beq
r_Q = 1-\frac{N_c}{N_f}, \quad
r_M = 2 r_Q = 2 - 2 \frac{N_c}{N_f}, \quad
r_q = \frac{1}{2} (2 - r_M) = \frac{N_c}{N_f}
\eeq
and so, due to the relation \eqref{rdim}, the anomalous dimensions obey
\beq
1-\gamma_Q = 3\frac{N_c}{N_f}, \quad \quad
1-\gamma_M = \frac{6 N_c - 3 N_f}{N_f}, \quad \quad
1-\gamma_q =3 \frac{N_f-N_c}{N_f}.
\eeq
So one has
\beq
(1-\gamma_Q) = (1-\gamma_M) + (1-\gamma_q).
\eeq
This relation is indeed consistent with \eqref{adler} as long as we replace the $q_f^2$ with $\Tr[T_F
T_F]$, the ``quadratic index" for the diagonal $SU(N_f)$ flavor symmetry with generators $T_F$. This index
is the appropriate non-Abelian generalization of $q_f^2$ and reduces to $q_f^2$ for any Abelian subgroup of the flavor
symmetry. The index is 1/2 for the $SU(N_f)$ fundamentals $Q$, $\tilde{Q}$, $q$, and $\tilde{q}$ and $N_f$ for the
$SU(N_f)$ adjoint meson field $M$. That is $N_f$ $Q$ and $\tilde{Q}$ pairs contribute as much as the single $M$-field.

So the proposed formula indeed gives the same Adler function in both theories, but was this an independent check? It is in fact easy see that the right hand side of \eqref{adler} in a conformal theory is just $3/2 \cdot \Tr(T_F T_F
T_R)$,
the mixed triangle anomaly between two flavor and one R-current. Again,
this relies on the fact the the superconformal algebra links the
R-charge to the dimension. But this triangle anomaly has to agree between the two theories
already by the standard 't Hooft anomaly matching arguments, which had previously been checked for all dual pairs.
So instead of being a new check, the equivalence of the Adler functions is guaranteed by anomaly matching.

Beyond confirming that the right hand sides of \eqref{adler} agree in the two theories, we also learn from \eqref{adler} what the actual value of the Adler function (the coefficient of the logarithm in the current/current 2-point function) is in the two conformal theories. This is in fact also a well known statement that played a crucial role in the discussion of a-maximization. In a superconformal field theory the flavor current correlation function is indeed given by the 't Hooft anomalies via \cite{Anselmi:1997ys,Barnes:2005bw}
\beq \langle j_{\mu}(x) j_{\nu}(0) \rangle = \frac{\tau}{(2 \pi)^4} (\partial^2 \delta_{\mu \nu}
- \partial_{\mu} \partial_{\nu}) \frac{1}{(x-y)^{2 (d-2)}} \eeq
with
\beq
\tau = - 3\, \Tr\left( T_R T_F T_F\right)
\eeq
from which \eqref{adler} at the conformal point follows.
When trying to use holography to verify these results one finds that all these identities are of course also obeyed by the dual supergravity \cite{Tachikawa:2005tq,Barnes:2005bw}. The very special geometry of ${\cal N}=2$ supergravity forces the bulk gauge coupling
(the prefactor of $F^2$ in the bulk), which sets the current 2-pt function, to be given in terms of the Chern-Simons coupling, which encodes the anomaly.

So to summarize, in super-conformal field theories the identity \eqref{adler} can be shown to be identical to known relations between current correlators and 't Hooft anomalies. The real interest in \eqref{adler} therefore arises when we move away from the conformal limit.

\section{Non-conformal Theories and Holography}

Many strongly coupled 3+1 dimensional supersymmetric gauge theories have a holographic dual description in terms of a classical theory of gravity in 4+1 dimensions.
Given our results from the last section, it should be clear that in order to test \eqref{adler} holographically, we need a background solution dual to a theory with at least one dimensionful coupling turned on to move away from the conformal point at which \eqref{adler} is automatically fulfilled. The simplest way to introduce a scale into the gauge theory is to give some of the matter field a mass. Most understood examples of holographic RG flows start with adding supersymmetry preserving mass terms to some fields. For simplicity we chose to work with the D3/D7 system \cite{Karch:2002sh}, which is the holographic dual description of adding $N_f$ ${\cal N}=2$ supersymmetry preserving hypermultiplets of mass $m$ to ${\cal N}=4$ SYM theory with $N_c$ colors and 't Hooft coupling $\lambda$. Due to the presence of the mass term this theory has scale dependence built in and so \eqref{adler} has non-trivial content.

Unfortunately, in the derivation of \eqref{adler} in \cite{Shifman:2014cya,Shifman:2015doa} set all mass terms are set to zero from the outset. It has not been made clear in that work whether this was just a simplifying assumption and \eqref{adler} was meant to hold more generally, or whether the validity of \eqref{adler} hinges entirely on the assumption that all matter multiplets are massless. We will see in our holographic example that \eqref{adler} would require a highly non-trivial structure for the anomalous dimensions in a theory with massive flavor, making it more likely that \eqref{adler} simply doesn't hold in theories with non-zero masses for the matter multiplets. We will confirm this interpretation in a free theory, where the anomalous dimensions vanish identically but the current correlator retains non-trivial momentum dependence.

\subsection{The holographic current correlation function for the D3/D7 system}

The left hand side of \eqref{adler} is a correlation function of gauge invariant operators, which makes it easy to compute using holography. The current operator maps to the worldvolume gauge field on the D7 and so we can read off the current/current correlation function of the field theory from the gauge field propagator in the bulk, following the standard dictionary.

This non-normalizable mode has e.g been found in
\cite{Hong:2003jm}.
Writing the induced AdS$_5$ $\times$ $S^5$ metric as
\beq
ds^2 = \frac{r^2}{R^2} \eta_{\mu \nu} dx^{\mu} dx^{\nu} + \frac{R^2}{r^2} dy_i dy^i
\eeq
with $\mu=0,1,2,3$, $i=4,\ldots,9$, and $r^2=y^i y_i$ the flavor brane worldvolume for a massive flavor multiplet of mass $m$ is is simply given by $y_8 = m \alpha' \equiv L$, $y_9=0$ so that the induced metric on the brane becomes
\beq
ds^2 = \frac{L^2}{R^2} (\rho^2+1) \eta_{\mu \nu} dx^{\mu} dx^{\nu}
+ \frac{R^2}{\rho^2+1} d \rho^2 + R^2 \frac{\rho^2}{\rho^2+1} d \Omega_3^2
\eeq
where we have introduced $\rho^2 = r^2/L^2-1$.
We have for the mode of interest for the gauge field
\beq
A_{\rho}=0, \quad A_{\beta}=0, \quad A_{\mu} = \xi_{\mu} \phi(\rho) e^{i k \cdot x}, \quad k \cdot \xi =0
\eeq
where $\beta$ labels the angles on the $S^3$,
\beq
\phi(w;\alpha) = \frac{\pi \alpha (1+\alpha)}{\sin(\pi \alpha)} F(- \alpha, 1+\alpha;2;w),
\eeq
$\rho^2/L^2=w/(1-w)$ and $\alpha=\frac{1}{2} (-1 + \sqrt{1- Q^2 \lambda/m^2})$ and hence
\beq \alpha (1+\alpha) = -\frac{Q^2 \lambda}{4 m^2}  .\eeq
This solution satisfies worldvolume equations of motion for small fluctuations as first written down in \cite{Kruczenski:2003be} and goes to 1 at the boundary as appropriate for a non-normalizable mode that we need to determine a current/current correlation function. From this non-normalizable solution we can simply extract the correlation function as \cite{Erlich:2005qh}
\beq
\Pi(Q^2) = - \frac{1}{g_5^2 Q^2} \left . \frac{\partial_z \phi(z)}{z} \right |_{z=\epsilon}
\eeq
where $z=r^{-1}$. $g_5^2$ is the effective 5d gauge coupling constant for the Maxwell theory on the D7 brane and is given
by\footnote{We normalize the gauge field so that the worldvolume action for the D7 brane takes its standard Dirac Born Infeld form,
\beq
S=-N_f T_{D7} \int d^8\xi \, \sqrt{\det(g_{ab} + (2 \pi \alpha') F_{ab})}.
\eeq
In this normalization, the endpoint of a string and hence a single quark carries charge 1 under the gauge field.}
\beq
g_5^{-2} = \frac{N_f N_c}{(2 \pi)^2  }.
\eeq
The resulting correlation is
\beq
\label{result}
\Pi(Q^2) = \frac{ N_f N_c }{8 \pi^2} \left ( H_{1+\alpha} + H_{- \alpha} + const. \right ) \eeq
where $H_x$ is the harmonic number,
\beq
H_n = \int_0^1 \frac{1-x^n}{1-x} dx .
\eeq
In order to obtain this correlation function we had to subtract a log divergent counterterm proportional to $\log(\epsilon)$ as is standard in holographic renormalization. There is however a contact term ambiguity in this procedure, leaving us with an undetermined constant in the correlation function.

The spectrum of this theory is a sequence of massive mesons, which are stable at large $N$. So we expect the spectral function of the theory to be given by a sequence of delta functions, localized at the masses of the mesons. That is the correlation function should be simply a sum over poles with the pole locations reproducing the known meson spectrum \cite{Kruczenski:2003be}. To see this, it is convenient to re-express the Harmonic numbers in terms of Digammma functions,
\beq
\psi(x) = \frac{\Gamma'(x)}{\Gamma(x)}
\eeq
via the identity
\beq
H_x = \psi(x+1) + \gamma
\eeq
where $\gamma$ is the Euler-Mascheroni constant, which can be absorbed in the undetermined constant. We obtain
\beq \Pi(Q^2) = \frac{ N_f N_c \lambda}{8 \pi^2 m^2} \left ( \psi(2+\alpha) + \psi(1- \alpha) + const. \right ). \eeq
The appearance of Digamma functions is somewhat reassuring. One of the easiest holographic current correlation functions is obtained in the soft-wall model \cite{Karch:2006pv}, whose spectrum is engineered to reproduce the phenomenology of Regge trajectories, that is mesons labeled by an integer $n$ with masses $m_n^2 \sim n$. In this case the correlation function is given
by \cite{Zuo:2008re} $\psi(1+q^2)$, which indeed yields the desired Regge spectrum as its poles.

In our case, we have the sum of two Digamma functions and so the situation is slightly more complicated. The spectrum of course is also not Regge like, but has masses grow with $n$, not $\sqrt{n}$ \cite{Kruczenski:2003be}. To proceed, we use the representation of the Digamma as an infinite sum:
\beq \psi(x) = - \gamma + \sum_{n=0}^{\infty} \left ( \frac{1}{n+1} - \frac{1}{n+x} \right  ) . \eeq
In this formula the individual sums diverge, as they go (at large $n$) as $\sum n^{-1}$. But the divergence cancels between the two terms. Since we have the freedom to drop (even infinite) constants from $\Pi(Q^2)$ we can drop the first term in the sum (which is $x$ and hence $\alpha$-independent). This helps to easily identify the meson spectrum:
\begin{eqnarray} \nonumber
\Pi(Q^2) &\sim& \sum_{n=0}^{\infty} \left ( \frac{1}{n+1-\alpha} + \frac{1}{n+2+\alpha} \right )
    \\  \nonumber &=& \sum_{n=1}^{\infty} \left ( \frac{1}{n-\alpha} + \frac{1}{n+1+\alpha} \right )
   \\  &=& \sum_{n=1}^{\infty} \frac{2n+1}{n(n+1) - \alpha (\alpha+1)}.
\end{eqnarray}
Last but not least, we plug back in  $\alpha(\alpha+1) = -Q^2 \lambda/(4m^2)$ to obtain
\beq \Pi(Q^2) \sim 4 \sum_{n=1}^{\infty} \frac{2n+1}{4 n(n+1) + Q^2}  + const. \eeq
The meson poles are at $-Q^2=M^2 =4 n (n+1) m^2/\lambda$, with $n=1,2,3,\ldots$. This is exactly where \cite{Kruczenski:2003be} tells us they should be in their equation (3.39). So this does indeed seem to be the correct correlation function.

\subsection{The anomalous dimensions}

In the previous subsection we calculate the left hand side of \eqref{adler} for the D3/D7 system. The resulting expression is highly non-trivial. The right hand side unfortunately is not as straightforward to obtain. The anomalous dimension of the fundamental gauge variant fields is not directly a gauge invariant observable, and hence not anything holography has directly access to. We need to rely on more indirect means in order to be able to deduce what the anomalous dimensions are in this setup. In fact, one could read our result \eqref{result} as a prediction for the anomalous dimensions of the flavor fields at strong coupling.

One limit in which we take $Q^2$ to be large and positive (that is spacelike momentum), we can expand $\Pi(Q^2)$ as
\beq \label{pt} \Pi(Q^2)=
-\frac{ N_f N_c }{8 \pi^2 } \left [  \log (Q^2) +  2 ( \gamma - \log 2) \right ] + {\cal O}(1/Q^2)
\eeq
The leading term gives rise to an Adler function of $D=3/2 N_f N_c$, which corresponds to a vanishing anomalous dimension according to \eqref{adler}. This appears consistent with expectations. Our theory has ${\cal N}=2$ supersymmetry and we do not expect any perturbative contributions to the anomalous dimensions. From this point of view the $Q^{-2}$ contributions to the Adler function look puzzling. They potentially can be interpreted as non-perturbative corrections to the anomalous dimensions. Alternatively, the relation \eqref{adler} may simply not hold in theories with massive quarks.

\subsection{Beyond a probe approximation in the presence of quark masses}

In this subsection, we extend the results in the preceding
subsection by studying a fully-backreacted D7-branes
in AdS$_5\times S^5$.
A first attempt along these lines was made in
\cite{Aharony:1998xz, Kehagias:1998gn}, which
computed the leading contribution to the fully-backreacted metric
of D3-branes in 7-brane backgrounds of F-theory.
In the case of AdS$_5\times S^5$ with $N_f$ D7-branes, analyses
of the backreacted solutions of SUGRA have been made extensively
so far. The paper \cite{Burrington:2004id} formulated a set of the
equations of motion of IIB SUGRA on the basis of the metric ansatz
adopted in \cite{Aharony:1998xz, Kehagias:1998gn},
and obtained the local behavior of the fully-backreacted
solution near the D7-branes.
A key ingredient there is to utilize the fact that the complexified
dilaton $\tau=C_0+ie^{-\phi}$ must be a holomorphic function of the complex $z=y_8+iy_9$ because of a BPS condition.
See also \cite{Grana:2001xn} for more on the holomorphic nature
of $\tau$ and \cite{Kirsch:2005uy} for a more general backreacted solution away from the D7s.
Here $C_0$ is an RR-scalar and $y_8$, $y_9$ are
the transverse coordinates of the coincident $N_f$ D7-branes.
Then, the monodromy condition of $\tau$ that ensures $N_f$
units of the $C_0$ magnetic flux implies that $\tau(z)$ behaves
near the D7-branes as
\begin{align}
  \tau(z)=\frac{N_f}{2\pi i}\log(z-z_0) \ .
\end{align}
Here $z_0$ denotes the location of the D7-branes, being
proportional to a complex quark mass $m$.
By identifying $\tau$ with the complexified gauge
coupling as
\begin{align}
 \frac{\theta}{2\pi}+i\frac{8\pi^2}{g_{\rm YM}^2}
=-iN_f\log(z-z_0) \ ,
\end{align}
and furthermore assuming that
$z$ probes an RG scale
$m\ll\mu\ll\Lambda$,
we see that this equals a one-loop running gauge coupling.
Here, $\Lambda$ is a Landau pole
of the ${\mathcal N}=2$ gauge theory with $N_f$ flavors under
consideration.

Note that this identification is fraught with subtleties. First of all, $z$ is a complexified RG scale and, especially when $z$ is close to $z_0$, its phase matters. What $z$ really parametrizes is the moduli space of vacua of a probe D3 brane, corresponding to breaking an $SU(N+1)$ gauge theory down to $SU(N) \times U(1)$ with $\tau(z)$ being the coupling constant of the $U(1)$ factor. Only when $z \gg z_0$ can this unambiguously be identified with the RG scale. However the simple log solution is only valid in the vicinity of the D7. But if we take the interpretation of $z$ as an RG scale at face value,
the NSVZ $\beta$-function (\ref{beta}) then shows that
the matter fields gain no anomalous dimensions $\gamma=0$.
The formula (\ref{adler}) thus proposes
the current/current correlator to be given as
\begin{align}
 D(Q^2)=\frac{3}{2}N_cN_f \ .
\end{align}
As argued in \S 3.2, this result is in accord with the large $Q^2$
limit of the current/current correlator that was obtained by summing
up an infinite number of single meson exchanges.

\section{Free theories}

In order to see whether \eqref{adler} holds in theories with massive quarks, one easy place to check is the theory of 2 massive chiral multiplets $Q$ and $\tilde{Q}$. They get mass via a superpotential
\beq
W = m Q \tilde{Q}.
\eeq
In components, this theory has 2 complex scalars $\phi$ and $\tilde{\phi}$ and a single Dirac fermion $\psi$ (since each chiral multiplet contributes a Weyl fermion).
This theory has a global $U(1)$ symmetry under which $Q$ has charge $+1$
and $\tilde{Q}$ charge $-1$. The corresponding current is given by
\beq
j^{\mu} = \bar{\psi} \gamma^{\mu} \psi - i(\phi^* \partial^{\mu} \phi - \phi \partial^{\mu} \phi^*)
+ i (\tilde{\phi}^* \partial^{\mu} \tilde{\phi} - \tilde{\phi} \partial^{\mu} \tilde{\phi}^*).
\eeq
In this case the anomalous dimensions vanish identically since the theory is non-interacting and so \eqref{adler} predicts that the current correlator has to reproduce the pure $\log(Q^2)$ behavior of a conformal theory. This sounds highly implausible as the correlator should surely depend on the value $m$ of the mass, but it will be good to get the explicit answer in any case. The correlation function simply is the sum of 3 individual pieces since $\psi$, $\phi$ and $\tilde{\phi}$ don't mix. Furthermore, the contribution from $\phi$ and $\tilde{\phi}$ is identical, since the extra sign squares to +1 in the 2-pt function. So we only have to calculate two diagrams, one for $\phi$ and one for $\psi$.

Let us start with the fermion. The relevant Feynman diagram simply has the two current insertions connected by free fermion propagators. In fact, this is really just the 1-loop contribution to the photon self-energy in standard spinor QED. For the current correlator in the free theory this one-loop diagram is the exact answer for the correlation function. It yields for
\beq
iA^{\mu \nu}(q) \equiv i \int d^4x \, e^{i qx} \, \langle T ( \bar{\psi} \gamma^{\mu} \psi(x) \bar{\psi} \gamma^{\mu} \psi(0) \rangle
\eeq
the following integral\footnote{Our conventions in here follow \cite{Srednicki:2007qs}. In particular, we use a mostly plus metric and the convention for the $\gamma$ matrices is $\{ \gamma^{\mu}, \gamma^{\nu} \} = - 2 g^{\mu \nu}$. This yields for the traces
$$ \Tr \gamma^{\mu} \gamma^{\nu} = - 4 g^{\mu \nu}, \quad \Tr \gamma^{\mu} \gamma^{\nu} \gamma^{\rho} \gamma^{\sigma} =
4 g^{\mu \nu} g^{\rho \sigma} - 4 g^{\mu \rho} g^{\nu \sigma} + 4 g^{\mu \sigma} g^{\nu \rho}.$$
{}Furthermore, $q^2$ in this section is identical to
$Q^2$ that is used in the definition of the Adler function
in section 1.}
\begin{eqnarray}
i A^{\mu \nu}(q) &=& - \int \frac{d^4k}{(2\pi)^4} \Tr \left ( \gamma^{\mu} \frac{-(\slashed{k} + \slashed{q})+m }{(k+q)^2 + m^2 - i
\epsilon} \gamma^{\nu} \frac{ -\slashed{k} +m}{k^2+m^2 - i \epsilon} \right ) \nonumber \\
&=& 4 \int \frac{d^4k}{(2\pi)^4} \, \frac{g^{\mu \nu}[ m^2+(k+q) \cdot k] -(k+q)^{\mu} k^{\nu} -(k+q)^{\nu} k^{\mu} }
{[(k+q)^2+m^2 - i \epsilon] \, [ k^2 +m^2 - i \epsilon]}
\end{eqnarray}
This integral is UV divergent as it contains infinite contact terms.
Using dimensional regularization gives
\begin{align}
  iA^{\mu\nu}(q)=4\int_0^1 dx\int \frac{d^dk}{(2\pi)^d}
\frac{\left(1-\frac{2}{d}\right)g^{\mu\nu}\,k^2
+g^{\mu\nu}\left(
-x(1-x)q^2+m^2\right)
+2x(1-x)q^\mu q^\nu}
{\left(k^2+x(1-x)q^2+m^2-i\epsilon\right)^2}\ .
\end{align}

For the scalar field, we can also follow the standard textbook treatment of scalar QED, see e.g section 65 of \cite{Srednicki:2007qs}. In this case two diagrams contribute to the photon self-energy, the single scalar bubble connected to two separate external photon lines that can directly interpreted as the current/current 2-pt function, as well as the scalar bubble with a single scalar-scalar-photon-photon vertex. Latter can be interpreted as a contact term contributing to the 2-pt function of the currents. Without it, the current correlator doesn't have the $q^{\mu} q^{\nu} - q^2$ demanded by the Ward identity implied by current conservation. Latter only holds up to contact terms, so we should not be surprised that we need to add two diagrams. Then,
\begin{align}
iB^{\mu \nu}(q) &\equiv -i \int d^4x \, e^{i qx} \, \langle T ( (\phi^* \partial^{\mu} \phi - \phi \partial^{\mu} \phi^*)(x) (\phi^* \partial^{\mu} \phi - \phi \partial^{\mu} \phi^*)(0) \rangle
\nonumber\\
&=
\int_0^1 dx\int \frac{d^dk}{(2\pi)^d}
\frac{-2\left(1-\frac{2}{d}\right)g^{\mu\nu}\,k^2
-2g^{\mu\nu}\left(
(1-x)^2q^2+m^2\right)
+(1-2x)^2q^\mu q^\nu}
{\left(k^2+x(1-x)q^2+m^2-i\epsilon\right)^2}\ .
\end{align}

Putting all together, we obtain
\begin{align}
  i\Pi^{\mu\nu}(q)=&i(A^{\mu\nu}(q)+2B^{\mu\nu}(q))
\nonumber\\
=&
\int_0^1 dx\int \frac{d^dk}{(2\pi)^d}
\frac{-4(1-x)g^{\mu\nu}q^2
+2q^\mu q^\nu}
{\left(k^2+x(1-x)q^2+m^2-i\epsilon\right)^2}
\nonumber\\
=&
i(q^2 g^{\mu \nu} - q^{\mu} q^{\nu} ) \Pi(q^2)  \ ,
\end{align}
with
%
%
\begin{align}
\Pi(q^2) = -\frac{1}{4\pi^2\varepsilon}
+\frac{1}{8 \pi^2} \int_0^1 dx \,  \log \left (
\frac{x(1-x)q^2+m^2}{\mu^2} \right ) \ .
\end{align}
Here $\varepsilon=4-d$.
We neglect the divergent term hereafter.
While much simpler in structure than the individual boson and fermion contribution, this integral of course nevertheless gives some interesting $m$ and $q$ dependent function, that is definitely in disagreement with \eqref{adler}:
\begin{align}
\Pi(q^2) = \frac{1}{4 \pi^2} \sqrt{1 +\frac{4 m^2}{q^2}} \mathrm{arccsch}
\left( 2 \sqrt{\frac{m^2}{q^2}} \right)
+ \frac{1}{8\pi^2}\log\frac{m^2}{e^2\mu^2} \ .
\end{align}
Clearly \eqref{adler} just doesn't hold for theories with massive matter multiplets.

\section{Discussion and Future Direction}

We found that the relation \eqref{adler} is automatically true in superconformal theories and fails to hold in theories with massive matter. What remains to be seen is under what circumstances
the theorem of \cite{Shifman:2014cya,Shifman:2015doa} contains some interesting new information. One obvious question one would like to address is whether \eqref{adler} allows for some non-trivial generalization to the case of massive matter. Our two examples, the free theory as well as the holographic D3/D7 system show that any such generalization would have to allow to reproduce highly non-trivial structure in the correlation function from more or less trivial anomalous dimensions.

Another question would be whether we can apply \eqref{adler} in holographic RG flows dual to gauge theories with massless matter.
We would like to have a theory with a running coupling constant but no mass terms. One thing one might hope to learn from such an example is how to identify the preferred scheme in which \eqref{adler} and \eqref{beta} on the gravity side. Current/current correlation functions in a supersymmetric flow with only ${\cal N}=1$ supersymmetry have been worked out for example in \cite{Bianchi:2001kw}. Unfortunately, this flow was triggered once again by a supersymmetric mass term, so presumably \eqref{adler} does not apply in this case either. This seems to be a genuine trap the holographic approach runs into. If we want to study a RG flow with a well understood UV, we need to start with a conformal field theory in the UV where conformal invariance is broken by turning on a relevant perturbation. This however seems to be exactly the situation in which \eqref{adler} does not apply.

\section*{Acknowledgements}

The work of AK was supported in part by the US Department of Energy under grant number DE-SC0011637 and a JSPS Invitation Fellowship for Research in Japan(Short-term) No.S-15031. AK wishes to thank the KMI at Nagoya University for their continued hospitality which made this collaboration possible.

\bibliographystyle{JHEP}
\bibliography{adler3}

\end{document}